\documentclass[useAMS,usenatbib]{mn2e}

\usepackage{graphicx}
\usepackage{threeparttable}
\usepackage{txfonts}
\usepackage{comment}
\usepackage{color}
\usepackage{natbib}

\title[Mid-infrared spectroscopy of SVS13]{Mid-infrared
spectroscopy of SVS13: Silicates, quartz and SiC in a
protoplanetary disc\thanks{
Based on data collected at Subaru Telescope, which is
operated by the National Astronomical Observatory of
Japan.
}}
\author[Fujiyoshi et al.]{Takuya
Fujiyoshi$^{1}$\thanks{E-mail: tak@subaru.naoj.org},
Christopher M. Wright$^{2}$ \& Toby J. T. Moore$^{3}$\\
$^{1}$Subaru Telescope,
National Astronomical Observatory of Japan,
National Institutes of Natural Sciences,
650 North A'ohoku Place, Hilo, HI 96720, USA\\ $^{2}$School
of Physical, Environmental and Mathematical Sciences,
UNSW Canberra, Canberra ACT 2600, Australia\\
$^{3}$Astrophysics Research Institute, Liverpool John
Moores University, IC2, Liverpool Science Park,
146 Brownlow Hill, Liverpool L3 5RF}
\begin{document}

\date{Accepted 2014 ??? ??. Received 2014 ??? ??; in
original form 2014 ??? ??}

\pagerange{\pageref{firstpage}--\pageref{lastpage}}
\pubyear{2014}

\maketitle

\label{firstpage}

\begin{abstract}
We present $N$-band (8$-$13~$\umu$m) spectroscopic
observations of the low-mass, embedded pre-main-sequence
close binary system SVS13. Absorption features are
clearly detected which are attributable to amorphous
silicates, crystalline forsterite, crystalline enstatite
and annealed SiO$_{2}$. Most intriguingly, a major
component of the dust in the envelope or disc around SVS13
appears to be SiC, required to model adequately both the
total intensity and polarisation spectra. Silicon carbide
is a species previously detected only in the spectra of
C-rich evolved star atmospheres, wherein it is a dust
condensate. It has not been unambiguously identified in
the interstellar medium, and never before in a molecular
cloud, let alone in close proximity to a forming star.
Yet pre-Solar grains of SiC have been identified in
meteorites, possibly suggesting an interesting parallel
between SVS13 and our own Solar-System evolution.
The uniqueness of the spectrum suggests that we are either
catching SVS13 in a short-lived evolutionary phase and/or
that there is something special about SVS13 itself that
makes it rare amongst young stars. We speculate on the
physical origin of the respective dust species and why
they are all simultaneously present toward SVS13. Two
scenarios are presented: a disc-instability-induced
fragmentation, with subsequent localised heating and
orbital evolution firstly annealing initially amorphous
silicates and then dispersing their crystalline products
throughout a circumstellar disc; and a newly discovered
shock-heating mechanism at the interface between the
circumstellar and circumbinary discs providing the
crystallisation process. One or both of these mechanisms
acting on carbon-rich grain material can also feasibly
produce the SiC signature.
\end{abstract}


\section{Introduction}

The Nebular Hypothesis, first formulated separately in
the 18th century by Swedenborg, Kant and Laplace, has
stood the test of time and is the generally favoured
overall description of how the Solar System formed and
evolved. In the young hot gaseous nebula,
micron-sized dust grains are dragged by the gas, and when
they collide they stick with one another and grow in size
by forming fluffy or porous aggregates (e.g.,
\citealt{Weidenschilling2000fop}; \citealt{Poppe2003soh}).
Although the details are still being worked out (see e.g.,
\citealt{Chiang2010fpi}), those aggregates continue to
grow by collision until they become $\ga$~1~km in size (now
called planetesimals), at which stage gravity takes
over and finally `runaway' growth of a few bodies leads to
planet formation (\citealt{Weidenschilling2000fop};
\citealt{Chiang2010fpi}). It is therefore of great
interest to study the mineralogy of circumstellar dust
around young stars as it represents the original
constituents of planetesimals and hence of the rocky
planets like our own Earth.

SVS13 (star number 13 in \citealt*{Strom1976iso})
is a pre-main-sequence (PMS) star \citep{Aspin2003tes}
thought to be driving the Herbig-Haro objects
7$-$11 \citep{Bachiller2000too} near the reflection
nebula NGC1333 in the Perseus molecular cloud. The
bolometric luminosity is estimated to be
$\sim$52~$L_{\sun}$ (\citealt*{Molinari1993tes};
their estimate of 115~$L_{\sun}$ at 350~pc scaled to an
updated distance of $235\pm18$~pc,
\citealt{Horita2008aoh}). Its optical and infrared
brightnesses increased considerably sometime between 1988
December and 1990 September \citep{Eisloeffel1991aoo}. The
brightening was wavelength dependent, with the object
becoming bluer as it underwent the change.
Even after this period of relatively large increase in
brightness ($\Delta m_{V} \sim 3.3$~mag, $\Delta m_{K}
\sim 1.2$~mag, \citealt{Eisloeffel1991aoo}), SVS13
showed a quasi-periodic near-infrared (NIR) brightness
fluctuation with amplitude $\sim$0.5~mag and cycle
$\sim$500~days between 1990 October and 1993 December
\citep{Aspin1994nmo}.

The 2.3-$\umu$m CO overtone band-heads, indicative of
the presence of hot ($\sim$3,000~K) and dense
($\geq$~10$^{10}$~cm$^{-3}$) molecular gas
\citep*{Carr2004hhe}, probably in the shape of a
circumstellar disc, have been observed in emission
(\citealt{Carr1989nce}; \citealt{Eisloeffel1991aoo};
\citealt{Biscaya1997fcv}; \citealt{Carr2004hhe}). The CO
band emission also displayed variability on timescales as
short as days \citep{Biscaya1997fcv}.

SVS13 is associated with a centimetre source
(VLA~4, \citealt*{Rodriguez1997is1}) which has been
resolved into a close binary (VLA~4A \& VLA~4B) separated
by $\sim$0.3~arcsec, first at 3.6~cm
\citep*{Anglada2000doa} then at 7~mm
\citep{Anglada2004asc}.
Although VLA~4A and 4B exhibit similar
flux densities at centimetre wavelengths, VLA~4B is the
dominant object (by a factor of more than 2) in the
millimetre range with its spectral index at these radio
wavelengths most probably arising from thermal
dust emission \citep{Anglada2004asc}.
The optical/NIR source SVS13 has been associated with
VLA~4A by these authors based on older optical data;
however, recently \citet{Hodapp2014tlr} identified SVS13
with VLA~4B, this time using the newer and more
(both photometrically and astrometrically) accurate 2MASS
catalogue (also, the epoch of the 2MASS observations
happens to fall between the epochs of the two radio
measurements that discovered and then confirmed the
close binary).
The separation ($\sim$0.3~arcsec), at the assumed distance
to the object of 235~pc, translates to 71~au, or slightly
outside the Kuiper Belt in our Solar System.

The first published mid-infrared (MIR) photometry of
SVS13 was conducted pre-outburst (\citealt{Cohen1983tes};
\citealt*{Harvey1984ioo}). A decade later, without
the knowledge of the brightening, \citet*{Liseau1992tiv}
obtained 10-$\umu$m measurements but did not find the
difference (from 10 years before) substantial. The first
$N$-band (8$-$13~$\umu$m) spectroscopy (in fact,
spectro-polarimetry) of the object was performed by
\citet{Aitken1993sim}, using the UCL spectro-polarimeter
on the 3.8-m UKIRT with an effective beam diameter of
4.3~arcsec and spectral resolution $R \sim 40$.
They remarked that the polarisation spectrum of SVS13 is
`unusual', in that it peaked
between 11 and 12~$\umu$m, not at the characteristic
wavelength of 10.2~$\umu$m for silicate grains.
Nevertheless, they considered the polarisation to arise
from dichroic absorption, albeit by an unusual
composition.
\citet{Wright1999msa} re-analysed the same data-set from
\citet{Aitken1993sim}, publishing the `unusual' spectrum
for the first time (see also \citealt{Smith2000sim}).
They concluded that the double-trough-shaped intensity
spectrum and the unique polarisation spectrum can
only be satisfactorily modelled using inclusions of SiC in
an amorphous silicate matrix (this point will be discussed
in detail in \S\ref{pol.sec}).

In this paper, we present the results of new $N$-band
spectroscopic observations of SVS13. The dust mineralogy
in the circumstellar environment is investigated by
fitting various dust emissivities to the spectrum
(\S\ref{model}).
We also discuss dust components that comprise the best-fit
model (\S\ref{revisit}) and suggest their possible origins
(\S\ref{cry.mech}).

\section{Observations and data reduction}

Table~\ref{obs-log} summarises the observations
of SVS13 and the standard star HD20644.
All measurements were made using the MIR imaging
spectrometer COMICS \citep{Kataza2000ctc} at the
Cassegrain focus of the 8.2-m Subaru Telescope on
Mauna Kea, Hawaii.
The $N$-band low spectral resolution ($R \sim 250$) mode
of COMICS utilises two Raytheon 320$\times$240 Si:As IBC
arrays, one as a slit-viewer and the other spectrometer,
and both are cooled by a Sumitomo 4-K Gifford-McMahon-type
cryo-cooler but usually operate at around 7$\sim$8~K
because of the self-heating. We used a 2-pixel wide slit
to achieve the spectral resolution of $R \sim 250$ in the
$N$-band.
The pixel scale of the spectrometer in the spatial
dimension is 0.165~arcsec so the slit width translates to
0.33~arcsec, which is comparable to the
diffraction-limited image size at 10~$\umu$m at 8-m class
telescopes.

\begin{table*}
\caption{COMICS observation log of SVS13 and the standard
star HD20644. Integration is the total on-source integration
time in seconds, and PA is position angle of slit in
degrees, measured East of North.}\label{obs-log}
\begin{tabular}{@{}lccrrc@{}}
\hline
Date       & Object  & Band  & Integration & PA          & Airmass \\
(UT)       &         &       & (sec)       & ($^{\circ}$) & \\
\hline
\multicolumn{6}{c}{{\em Spectroscopy}} \\
2009/11/03 & SVS13   & $N$   & 994         & 90          & 1.224$\rightarrow$2.125 \\
           & HD20644 & $N$   &  65         &  0          & 1.072$\rightarrow$2.659 \\
\hline
\multicolumn{6}{c}{{\em Imaging}} \\
2009/11/03 & SVS13   & N8.8  & 401         & $-$         & 1.081$\rightarrow$1.092 \\
           &         & N11.7 & 401         & $-$         & 1.102$\rightarrow$1.113 \\
           &         & N12.4 & 401         & $-$         & 1.149$\rightarrow$1.163 \\
           & HD20644 & N8.8  &  22         & $-$         & 1.059 \\
           &         & N11.7 &  21         & $-$         & 1.038 \\
           &         & N12.4 &  21         & $-$         & 1.047 \\
\hline
\end{tabular}

\end{table*}

Data reduction was carried out using the
{\sc iraf}\footnote{{\sc iraf} is distributed by the
National Optical Astronomy Observatories, which are
operated by the Association of Universities for Research
in Astronomy, Inc., under cooperative agreement with
the National Science Foundation.} data reduction and
analysis package. The chop-subtracted 2-D spectra were
flat-fielded and then geometry distortion corrected.
Wavelength calibration was achieved using numerous
telluric emission lines
present in the $N$-band. For flux calibration and
telluric absorption correction, we also observed a
Cohen standard star HD20644 \citep{Cohen1999sic}. We
furthermore imaged both the target and standard in
three different bands ($\lambda_{0}$ = 8.8, 11.7, and
12.4~$\umu$m) to rectify possible slit efficiency
discrepancy. The images of SVS13 and HD20644, in terms
of their point-spread functions, were very similar to
each other at all three bands [i.e., SVS13 is
unresolved. Indeed, when observing in the spectroscopy
mode, the slit was set to run East-West (instrument PA =
90$^{\circ}$) in an unsuccessful attempt to resolve the
close binary].
In fact, both sets of images of the target and the
standard at longer wavelengths, i.e., at 11.7 and
12.4~$\umu$m, at which they are even less susceptible to
seeing because of its $\lambda^{-\frac{1}{5}}$ wavelength
dependence, show the first Airy disc, a clear sign of the
diffraction-limited condition. We therefore conclude that
only one star (VLA~4B) in the close binary system has any
MIR emission associated with it.
\citet{Hodapp2014tlr} could also only find a single
object in their laser-guided adaptive-optics assisted $H$-
and $K$-band images.
As such, in the final stages of the data reduction, three
pixels in the spatial direction were simply summed up
(i.e., $3 \times 0.165 \sim 0.5$~arcsec) to improve the
signal-to-noise ratio.

\section{Results and model fitting}

\subsection{Spectrum}

\begin{figure}
\includegraphics[width=84mm]{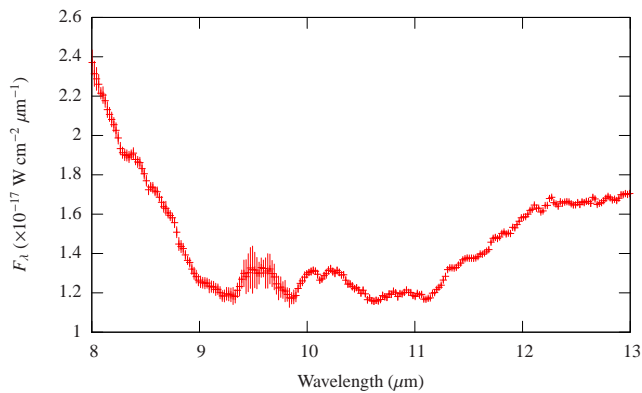}
\caption{COMICS $N$-band spectrum of SVS13. Note that
error-bars only represent standard deviation in the sky
background.}
\label{comics.spc}
\end{figure}

Figure~\ref{comics.spc} shows the COMICS $N$-band
spectrum of SVS13. Note that error-bars only represent
the standard deviation in the sky background and not
the uncertainty in the absolute flux calibration, which
would probably amount to about 10~per~cent (most of which,
approximately 4 to 7~per~cent, originates from the
uncertainty in the Cohen flux standard template at these
wavelengths).

The double-trough shape described in the Introduction is
clearly apparent; however, at the spectral resolution
$R \sim 250$, it remarkably becomes a triple-trough, with
a hump at around 9.5~$\umu$m.
The first general dip is most likely caused by the
silicate absorption, while the second, with its deepest
part at about 11~$\umu$m, could be due to the presence of
SiC, as argued by \citet{Wright1999msa}. The highest
spectral resolution so far attained on the object by
COMICS, compared to the relatively low resolution of UCLS
[$R \sim 40$, see fig.3(b) in \citet{Wright1999msa} and
fig.2.4 in \citet{Smith2000sim}], displays finer details
not previously seen, which will be utilised below for
model fitting.

\subsection{Model fitting}\label{model}

\subsubsection{Dust species}

Selecting dust species can be subjective exercise but is
constrained by those historically identified in
astronomical spectra or as (pre-Solar) components of
meteorites and interplanetary dust particles.
We therefore restrict ourselves in using the following
dust species and their combinations.

\begin{enumerate}

\item{\em Amorphous (astronomical) silicates}
(as represented by the
Trapezium emissivity, \citealt*{Forrest1975cga}).
Although their specific mineralogy has been extremely
difficult to identify, they
are the most abundant dust grains in the interstellar
medium (ISM; e.g., \citealt*{Molster2010tmo}).
The emissivity has been derived by dividing the observed
$N$-band spectrum of the Trapezium region
\citep{Forrest1975cga} by a 250-K Planck black-body
function \citep{Gillett1975t8m}.

\item{\em SiC} (carbon star emissivity,
\citealt{Aitken19798ms}).
It is a major component in carbon-rich stellar ejecta
(e.g., \citealt{Treffers1974hso}) and is also found in
meteorites as pre-Solar (probable extra-Solar-System)
grains (e.g., \citealt{Bernatowicz1987efi}); however, it
has never been observed in the ISM
\citep*{Whittet1990ota}. We note that, using an Infrared
Space Observatory ({\em ISO}) SWS spectrum, a later
study \citep{Min2007tsa} has found some evidence of the
existence of SiC grains in the $N$-band `ISM' absorption
feature in the line of sight towards the Galactic Centre
(GC) source Sgr~A$^{\ast}$; however, the large beam size
($14\times20$~arcsec$^{2}$) of {\em ISO} encompasses a
number of asymptotic giant stars (i.e., the most probable
source of meteoritic SiC grains, e.g.,
\citealt{Zinner1998sna}) identified in several catalogues
(\citealt*{Blum1996rcs}; \citealt*{Ott1999vae};
\citealt{Clenet2001aol}; \citealt{Blum2003rcs}).  We
therefore suspect it may be contaminated and so may not
represent the pure ISM extinction. \citet{Whittet1990ota}
studied $N$-band spectra of 10 GC sources from
\citet{Roche1985aio}, which were obtained using the UCL
spectrometer with a moderate beam diameter of 4.2 or
4.3~arcsec, thus those data should be less susceptible to
contamination by source confusion.
Regardless of its existence in the ISM, it has never been
detected in the circumstellar environment of young
stars.\footnote{
An over-abundance of carbon has been reported in the
debris disc of the archetypal young star $\beta$-Pictoris
\citep{Roberge2006sot}. Although SiC is not explicitly
referred to, it may be posited that one of the most likely
compounds which could form in such a carbon-rich
atmosphere is silicon carbide.
}
The emissivity has been derived by dividing the observed
$N$-band spectrum of the carbon star Y-Tau by a 800-K
Planck black-body function \citep{Aitken19798ms}.

\item{\em Crystalline silicates} (aerosol forsterite and
enstatite, \citealt{Tamanai2006t1m}). The {\em ISO}
`crystalline silicate revolution' revealed they are
frequently present in stars young and old
\citep{Jaeger1998sti}. Now, {\em Spitzer} has both
expanded and refined the crystalline silicate database
(e.g., \citealt{Olofsson2009css}); however, once again,
they have never been identified in the diffuse ISM
\citep*{Kemper2004toc} [a caveat here is that the work of
\citet{Min2007tsa} mentioned above used the same data-set
from \citet{Kemper2004toc}, i.e., the {\em ISO} SWS
spectrum of the GC source Sgr~A$^{\ast}$, which may not
describe the true nature of the ISM dust grains].
The two most abundant crystalline silicate species are
forsterite (Mg$_{2}$SiO$_{4}$) and enstatite (MgSiO$_{3}$,
\citealt{Molster2010tmo}). Commercially available
crystalline silicate dust powders were dispersed into a
nitrogen gas stream to create non-embedded, free-flying
particle samples for infrared extinction measurements
\citep{Tamanai2006t1m}.

\item{\em Amorphous and annealed SiO$_{2}$}
(\citealt{Fabian2000sti}). It is one of the most
abundant compounds in the Earth's crust but its features
have not been observed in the diffuse ISM (see e.g.,
\citealt{Li2002asn}). It has however been identified in a
number of protoplanetary discs around young stars
(e.g., \citealt{Sargent2009sip}). For the crystalline
component, we use the nano-particles of precipitated
silica annealed at 1,220~K for 5~hours
\citep{Fabian2000sti}.

\end{enumerate}

\begin{figure}
\includegraphics[width=84mm]{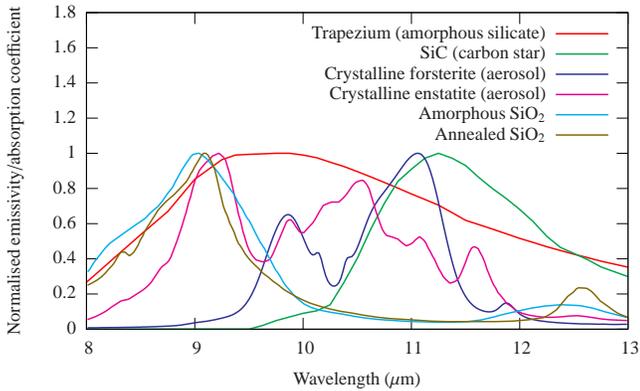}
\caption{Normalised emissivities/absorption coefficients
of the selected dust species (see text).}
\label{all-comp}
\end{figure}

We use either empirical (amorphous silicates and SiC) or
experimental (crystalline silicates and
amorphous/annealed SiO$_{2}$) results on an ensemble of
presumably differently shaped/sized grains. This approach
is slightly varied from a number of other mineralogical
studies which usually adopt some optical constants and
shape distributions to calculate absorption coefficients
(e.g., \citealt{Bohren1983aas}). It is
unrealistic to expect dust grains to have regular shapes
and indeed a sample of interplanetary particles consists
of irregularly shaped porous aggregates (see e.g.,
\citealt{Bradley2003idp}). Also, SVS13 is polarised in the
MIR (i.e., aligned aspherical grains must be
present, \citealt{Aitken1993sim}), hence we consider our
approach to be at least a more pragmatic one.
Normalised emissivities/absorption coefficients of the
materials used in this study are shown in
Figure~\ref{all-comp}.

\subsubsection{Model}

The fitting function can be expressed as
\[
F_{\lambda} \propto B_{\lambda}(T) \times e^{-\alpha}
\]
where $B_{\lambda}(T)$ is the Planck black-body
function at temperature $T$,  $\alpha =
\sum_{i=1}^k \tau_{i}$ is the total optical
depth of $k$ absorption components and $\tau_{i}$
is the optical depth of each component. In order
to keep the fitting as simple as possible, we did
not consider, for example, an extra black-body
component, self-absorption of a dust species, etc.

We used an {\sc iraf} task {\em specfit}
\citep{Kriss1994fmt} to fit the emissivities/absorption
coefficients. The fitting results are summarised in
Table~\ref{fit-results}. As the Planck black-body function
is always present, we distinguish between each model with
the number of absorption components ($k$-comp, where
$k$ is the number of different dust species) and the
names of specific absorptive ingredients added [am
and cr stand for amorphous and crystalline, and
Trap Trapezium (amorphous/astronomical silicates),
fors/fo forsterite, enst/en enstatite, respectively. Every
model is also given a unique alphabetical identification
index (A--I), as listed at the beginning of each row].

\begin{table*}
\caption{{\em specfit} results of various model fitting
(see text for details). In Column~1, alphabetical model
IDs and names of various models discussed in the text are
listed. The goodness of fit, $\chi^{2}_{\nu}$, is given in
Column~2. In Columns~3, the black-body temperature (in
$K$) is given. In Columns~4--15, optical depths of each
dust component and their associated uncertainties (given
by {\em specfit}, converted to per~cent) are
listed.}\label{fit-results}
\begin{tabular}{@{}ccr@{.}lccrcrcrcrcrcr@{}}
\hline
\multicolumn{2}{c}{Model}   & \multicolumn{2}{c}{$\chi^{2}_{\nu}$} & $T(K)$ & Trap & err    & SiC  & err   & Cr fo & err   & Cr en & err   & Am SiO$_{2}$ & err  & Cr SiO$_{2}$ & err \\
\hline
A & 1-comp                  & 30 & 0                              & 319    & 1.11 & 2\%    &      &       &       &       &       &       &             &       &             & \\
B & 2-comp                  & 10 & 5                              & 298    & 0.94 & 1\%    & 0.18 & 4\%   &       &       &       &       &             &       &             & \\
C & 2-comp+cr fors          &  8 & 33                             & 300    & 0.86 & 1\%    & 0.14 & 13\%  & 0.08  & 34\%  &       &       &             &       &             & \\
D & 2-comp+cr enst          &  6 & 51                             & 301    & 0.66 & 4\%    & 0.18 & 13\%  &       &       & 0.23  & 13\%  &             &       &             & \\
E & 2-comp+cr fo\&en        &  3 & 74                             & 302    & 0.55 & 2\%    & 0.15 & 4\%   & 0.10  & 5\%   & 0.24  & 1\%   &             &       &             & \\
F & 4-comp+am SiO$_{2}$     &  1 & 95                             & 313    & 0.56 & $<$1\% & 0.17 & 2\%   & 0.12  & 4\%   & 0.24  & 2\%   & 0.15        & 6\%   &             &\\
G & 4-comp+cr SiO$_{2}$     &  1 & 67                             & 309    & 0.60 & 1\%    & 0.18 & 3\%   & 0.11  & 5\%   & 0.21  & 5\%   &             &       & 0.16        & 2\% \\
H & 4-comp+am\&cr SiO$_{2}$ &  1 & 67                             & 310    & 0.59 & $<$1\% & 0.18 & 3\%   & 0.11  & 6\%   & 0.21  & 3\%   & 0.02        & 29\%  & 0.14       & 4\% \\
I & G$-$SiC                 & 14 & 7                              & 314    & 0.56 & $<$1\% &      &       & 0.18  & 4\%   & 0.23  & 4\%   &             &       &             &\\
\hline
\end{tabular}
\end{table*}

The goodness of fit is given by the reduced
chi-square, $\chi^{2}_{\nu} \equiv \chi^{2}/\nu$,
where $\nu$ is the number of degrees of freedom.
Since there is only one emissive component (a black-body)
and the rest are always absorptive, which all naturally
come after the black-body, the order in which the
absorptive components appear should not matter (i.e., they
are commutative). We have indeed obtained exactly the same
results (minimum $\chi^{2}_{\nu}$ and $\tau$ values) even
when the order in the component list was altered.
It is interesting to note that when the SiC ({\em plus}
Trapezium) emissivity is combined with only one of the
crystalline silicate species (forsterite or enstatite),
their best-fit parameter values are rather uncertain [by
more than 10~per~cent; see models C (2-comp+cr fors) and D
(2-comp+cr enst) in Table~\ref{fit-results}]; however,
when all three (SiC, crystalline forsterite and enstatite,
{\em plus} Trapezium) are combined, their best-fit values
become significantly more certain [by $\leq$~5~per~cent;
see model E (2-comp+cr fo\&en)]. This is most probably a
manifestation of the fact that it takes all three
emissivities/absorption coefficients to fit well the
complex spectral shape longward of $\sim$10~$\umu$m.

\begin{figure}
\includegraphics[width=84mm]{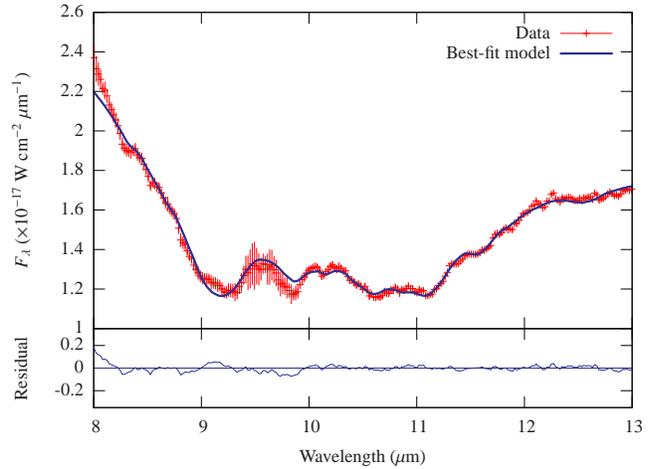}
\caption{{\em Upper panel}: $N$-band spectrum of SVS13
overlaid with the best-fit model (solid line). {\em Lower
panel}: Fitting residual plot.}
\label{best.fit}
\end{figure}

The general trend is clearly such that the more
components there are, the better the fit, as one
might expect. However, it stops
when both amorphous and annealed SiO$_{2}$ are added to
the mix. Were the trend to continue, one would have
expected to obtain the smallest $\chi^{2}_{\nu}$ for the
combination that involves all the dust species considered
in this study. Not only did $\chi^{2}_{\nu}$ not improve
but also the contribution from amorphous SiO$_{2}$
became insignificant (in addition, its best-fit
$\tau$ is poorly constrained, with the associated
uncertainty at nearly 30~per~cent, while all the other
values show reasonable certainty at well below
10~per~cent). Combining these findings with the fact that
adding annealed SiO$_{2}$ alone gives a better fit than
the amorphous counterpart in its place, we consider model
G (4-comp+cr SiO$_{2}$, i.e., a black-body, amorphous
silicates, SiC, crystalline forsterite and enstatite, plus
annealed SiO$_{2}$) the best-fit model with the fewest
number of components. Figure~\ref{best.fit} shows the
$N$-band spectrum of SVS13 overlaid with the best-fit
model (solid line).

The model's ability to reproduce fine structures,
especially the longward of $\sim$10~$\umu$m, is
remarkable, considering the complexity of the spectrum.
The only obvious shortcomings of the model
prediction are: the slightly shallower
slope towards the short-end of the $N$-band window;
not reproducing well the shape of the 8.3-$\umu$m
absorption feature;
the slightly deeper absorption at about 9.1~$\umu$m;
slightly shallower absorption at around 9.8~$\umu$m;
not predicting adequately the sharpness of the
10.1-$\umu$m absorption signature,
et~cetera (see also the fitting residual plot in the
lower panel of Figure~\ref{best.fit}).
However, it has been known
for some time that, for laboratory measured absorption
spectra of dust species, the method of preparation of
dust samples is critically important. For example,
hand-grinding and ball-milling will produce different
results \citep{Koike2010eof}. Annealing processes
(at different temperatures and for different durations)
also affect the outcome and depending on the treatment,
peak positions and full-widths at half-maximum of dust
features will differ, sometimes significantly
(\citealt{Koike2010eof}; see also
\citealt{Fabian2000sti}). When these further complications
are taken into account, it is even more impressive
how good a fit the selected model components provided,
considering the complexity of the features modelled and
the limited number of dust species required. In the
following discussion, we assume that the dust components
present in the best-fit model exist in the
circumstellar environment of the PMS star SVS13.

\section{Discussion}

\subsection{Dust species (revisited)}\label{revisit}

\subsubsection{Amorphous (astronomical) silicates}

As mentioned earlier, they are the most abundant
dust species in the diffuse ISM \citep{Molster2010tmo}.
The broad, smooth absorption feature arising
from the Si$-$O stretching mode, peaking at about
9.7~$\umu$m, is the strongest and best studied
infrared feature in the diffuse ISM (e.g.,
\citealt{Whittet2003dit}),
although identifying their specific mineralogy or
composition has traditionally been extremely difficult,
especially if accompanying 20-$\umu$m data is not
available.
Most, if not all, astronomical mineralogy
studies contain amorphous silicates in one form or
another (see, for example, tab.1 in
\citealt{Molster2010tmo}).
Therefore, it is not surprising that
the amorphous silicate absorption appears
prominently in the $N$-band spectrum of a young star.
However, we do not attempt to associate it with a
particular mineral, but rather use a generic
representation of amorphous silicates in a star-forming
region as provided by the Trapezium emissivity.

\subsubsection{SiC}\label{sic.sec}

Silicon carbide has never been observed in the diffuse
ISM \citep{Whittet1990ota}, although it is produced in
carbon star ejecta and is undoubtedly injected into it.
So where does it go?
One simple explanation may be that dust grains are
destroyed in the ISM by, for example, supernova shock
waves (e.g., \citealt*{Jones1996gsi}). If the substantial
gap between the dust lifetime ($\sim$10$^{8}$~yr,
\citealt{Jones1994gdi}; 1996\nocite{Jones1996gsi}) and the
dust injection timescale (from dust formation in the
outflows of dying stars to being incorporated into new
star-forming regions; $\sim$10$^{9}$~yr,
\citealt{Dwek1980teo}; \citealt{Jones1994idp}) is valid,
then SiC grains may also have been destroyed, hence no
11.3-$\umu$m SiC feature detected in the ISM.
However, as mentioned earlier, pre-Solar SiC grains
(and many others) have been found (e.g., in a carbonaceous
chondrite Murray by \citealt{Bernatowicz1987efi}) so some
dust grains must survive the harsh ISM.

SiC has in fact also been suggested as a dust component in
a number of comets \citep*{Orofino1994sca}. Good fits to
the $N$-band spectra of three comets (1P/Halley, C/1987 P1
Bradfield, and 1986l Wilson), which all exhibited two
broad peaks centred at about 9.7 and
11.3~$\umu$m,\footnote{Comet Wilson showed an extra,
unidentified emission feature at around 12.5~$\umu$m but
it was omitted from fitting procedure
\citep{Orofino1994sca}.} were obtained using a laboratory
measured spectrum of synthetic amorphous olivine (i.e.,
silicate, \citealt{Blanco1991lso}) and the complex
refractive index of $\alpha$-SiC from
\citet{Pegourie1988opo}.
It should be noted, however, that the 11.3-$\umu$m feature
in comets has generally been attributed to crystalline
silicates (e.g., for Halley's comet,
\citealt{Campins1989tio}). While the available data is
limited to the $N$-band spectrum alone, its identification
unfortunately remains rather ambiguous as their signatures
are often overwhelmed by that of amorphous silicates. Only
when measurements at other wavelengths (especially at
20~$\umu$m and longer) are on hand, the crystalline
silicate assignment becomes more robust (e.g., for Comet
Hale-Bopp using the {\it ISO} SWS data,
\citealt{Crovisier1996tis};
1997\nocite{Crovisier1997tso}).
Those three comets (and Hale-Bopp also) are all
long-period Oort Cloud comets which are thought to have
originated just beyond the ice giant formation region
($\ga$~15~au, \citealt{Charnoz2007cda};
\citealt{Levison2008oot}).

The most conclusive evidence for the presence of pre-Solar
SiC grains in comets came from dust samples, captured and
returned to Earth by NASA's Stardust mission to Comet
81P/Wild~2 \citep{Messenger2009dop}, which belongs to
short-period Jupiter Family of comets (JFCs), which are
likely to have formed in the Kuiper Belt ($\ga$~30~au).
The discovered 300-nm SiC grain is unique, in that it is
the only pre-Solar grain so far found in the silica
aerogel dust collector, possibly reflecting the complex
preparation required for the aerogel track analysis
\citep{Floss2013tao}, and no pre-Solar SiC grains have
yet been detected in the impact craters on the surface
of exposed aluminium foils. The abundance of pre-Solar
SiC grains in 81P/Wild~2 has been estimated to be
$\sim$45~ppm \citep{Floss2013tao}, which is consistent
with that in insoluble organic matter residues from a
number of different classes of chondritic meteorites
(10$-$55~ppm, \citealt{Davidson2009psa}).
These findings reveal that SiC grains must have been
available in a wide range of distances from the Sun, from
the carbonaceous chondrite formation region of the inner
few au to the comet-forming outer reaches of the
proto-Solar nebula.

Interestingly, pre-Solar SiC grains found in another
carbonaceous chondritic meteorite Murchison, for example,
showed little evidence of scarring
\citep{Bernatowicz2003pps} due, for instance, to
grain-grain collisions. Were the interstellar environment
as destructive as it has been described (e.g., by
\citealt{Jones1996gsi}), such hallmarks should certainly
be present. Then again, whether a dust grain survives is
probably largely statistical in nature and it is perhaps
only natural that those that do show little damage.
Nonetheless, when \citet{Jones2011ddi} re-examined the
dust lifetime using revised uncertainty estimates, they
found it to be comparable with the dust injection
timescale.

\citet{Whittet1990ota} proposed an alternative solution
to the missing SiC problem: oxidation. They argued that
the ejected SiC grains could be selectively destroyed by
surface oxidation in the O-rich interstellar environment.
Indeed, partial oxide layers on some pristine pre-Solar
SiC grains from the Murchison meteorite have been reported
\citep*{Croat2010sso}. In volatilisation
experiments of SiC grains under Solar nebula-like
oxidising conditions \citep{Mendybaev2002vko}, either
continuous or partial SiO$_{2}$ layers are formed,
depending on oxygen fugacities (continuous at higher
and partial at lower fugacities, or more and less
oxidising, respectively).
Similarly, though probably unintentionally, an oxide
layer had grown on the surface of SiC nano-particles
after they were taken out of a vacuum chamber
\citep{Clement2003nls}. However, even though a
SiO$_{2}$ layer does add new features due to the presence
of the mantle material, it does not appear to suppress very
much the 11.3-$\umu$m SiC feature
(\citealt*{Posch2004rto}; \citealt*{Zhang2009otc}).

In a similar study to \citet{Croat2010sso},
\citet*{Croat2009aan} and \citet*{Croat2010psc} identified
coatings mainly consisting of carbonaceous material on the
surface of some pristine pre-Solar SiC grains also from
Murchison. Furthermore, once again from the same
meteorite, some graphite spherules were found to contain
SiC grains (\citealt{Bernatowicz1996cos};
\citealt{Croat2006scw}; \citealt*{Hynes2007mos};
\citealt*{Croat2008cia}; \citealt{Croat2008esa};
\citealt*{Croat2010u2s}). 
Such SiC-containing graphites have also been found in
samples from another meteorite, Orgueil
(\citealt{Croat2009lpg}; \citealt*{Croat2014tso}). The
discoveries of these SiC-containing graphites are quite
significant as they are clear evidence that pre-Solar
SiC grains either formed earlier than or at about the same
time as the graphites. One spherule in particular (only
one of a kind so far), discovered by \citet{Croat2008esa},
houses a SiC grain at its centre, most probably indicating
the central SiC acted as a nucleation core (i.e., SiC may
have been formed first).
It should be noted, however, that the number of SiC
inclusions hitherto found is rather limited; only about a
dozen graphite spherules that contain SiC grains have been
reported (see references listed above), whilst graphites
with internal carbides (predominantly TiC,
\citealt{Croat2008cia}) number $\sim$80 to date, despite
the fact that the abundance of Ti is very much lower than
that of Si.

These results (the scarcity of SiC-containing graphites
and much more numerous TiC inclusions) can be explained by
a relative condensation sequence of solids in carbon star
outflows; in conditions typically found for carbon stars,
the condensation sequence is thought to be TiC, graphite,
and then SiC \citep{Lodders1995too}. There have however
been some suggestions that, under certain circumstances,
SiC is formed before graphite (\citealt{Chigai2003cso};
\citealt{Yasuda2012fos}).
In fact, condensation in this order (SiC before graphite)
has been invoked to account for the lack of the
11.3-$\umu$m SiC feature in the diffuse ISM
(\citealt*{Frenklach1989sca}; \citealt{Kozasa1996fos}). As
SiC grains formed in the vicinity of carbon-rich stars
travel further away and become cooler, a carbon layer
develops on the grains and hides the SiC feature. Such
scenario was in fact briefly mentioned by
\citet{Whittet1990ota}, and unlike the oxide coating
discussed earlier, the carbon mantle seems to suppress
successfully the 11.3-$\umu$m signature
(\citealt{Kozasa1996fos}; \citealt{Papoular2008tso}).

A short letter has been published in a surface science
journal that reports the dissolution of the amorphous
carbon layer into the SiC core, when the core-mantle grain
was heated to 600$\degr$C (873.15~K) under high vacuum
($3\times10^{-8}$~Torr, \citealt*{Kimura2003hbo}). At
800$\degr$C (1,073.15~K), the surface layer completely
disappeared without altering the lattice structure of the
particle.
Somewhat surprisingly, they have further observed that
when the SiC grain was cooled back to room temperature,
still under high vacuum and thus without any significant
contaminants, the amorphous carbon coating re-emerged
(i.e., the process is reversible,
\citealt{Kimura2003hbo}).
The temperature reached in their experiment (1,073.15~K)
is above the glass transition temperature of amorphous
analogues of enstatite (MgSO$_{3}$) composition (1,040~K,
\citealt{Roskosz2011asc}).
As crystalline dust species (crystalline
forsterite and enstatite, and annealed SiO$_{2}$;
see \S\ref{cry.sil} \& \ref{ann.sio2}) are also present,
temperatures reached in the circumstellar environment of
SVS13 must have been high enough to produce those and
remove the carbon coating from the SiC grains.
Under presumably much more severe conditions encountered,
the amorphous carbon layer could still be absent from the
SiC core even at sufficiently low temperatures
($\la$~100~K at $\ga$~15~au, \citealt{Poteet2011asi}) for
the grains to exhibit the SiC absorption feature. Recall
that \citet{Wright1999msa} concluded that SiC grains are
most likely to be included in an amorphous silicate matrix
(see also \S\ref{pol.sec}).

To our knowledge, it has never been established whether
the re-heated SiC particles would exhibit (in absorption)
exactly the same feature as that seen towards carbon
stars. In any event, it is known to vary somewhat from one
carbon star to another; so much so that,
\citet*{Speck2005teo} have coined a term `$\sim$11-$\umu$m
feature' (i.e., with the approximation mark `$\sim$' and
fewer significant figures) to emphasise its variability.
Also, a self-absorbed SiC band has been detected towards
extreme carbon stars (\citealt*{Speck1997tno};
\citealt{Justtanont1997iso}; \citealt*{Hony2002tco}),
where amorphous carbon might be expected to dominate.
Therefore, although this scenario (resurgence of the SiC
feature in a warm environment) may plausibly explain the
situation in SVS13, it should be treated with caution.

\begin{figure}
\includegraphics[width=84mm]{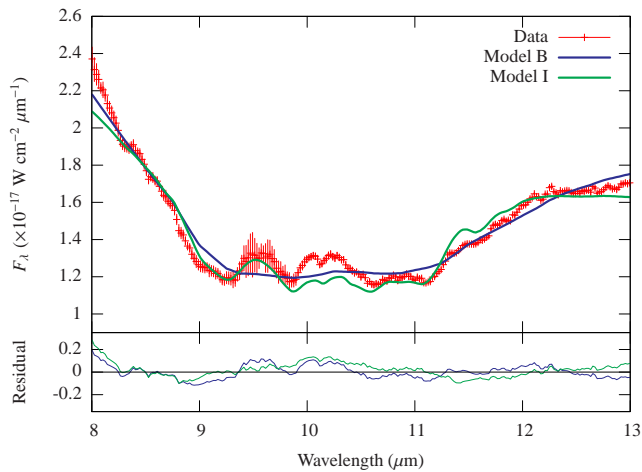}
\caption{{\em Upper panel}: $N$-band spectrum of SVS13
overlaid with models B and I. {\em Lower panel}: Fitting
residual plot for the two models.}
\label{bf-sic}
\end{figure}

However, we highlight the fact here that if we take out
the SiC component from the best-fit model
($\rm{I}\equiv\rm{G}-\rm{SiC}$ = a black-body, amorphous
silicates, crystalline forsterite and
enstatite),\footnote{When SiC was taken out of the
best-fit model, the contribution from annealed SiO$_{2}$
became rather insignificant or even negative (i.e., turned
emissive). We have therefore decided to exclude it from
the fitting procedure in this test case. The inference
drawn in this section would still be valid regardless of
the crystalline SiO$_{2}$ inclusion.}
the resulting fitting $\chi^{2}_{\nu}$ value (14.7) is
better than that for the most basic model A
(1-comp = a black-body plus amorphous silicates;
30.0) but worse even than the one for the simple model B
(2-comp = a black-body, amorphous silicates, and SiC;
10.5), signifying need for the SiC inclusion.
See Figure~\ref{bf-sic} which compares models~B and I. It
is clear, apparent in the residual plot in the
lower panel, that the fit for model~I becomes much worse
than that for model~B longward of $\sim$10~$\umu$m,
especially around 11.3~$\umu$m, where the contribution
from the SiC emissivity is significant.

\paragraph{Polarimetry perspective}\label{pol.sec}

\begin{figure*}
\includegraphics[width=1.0\textwidth, angle=180]{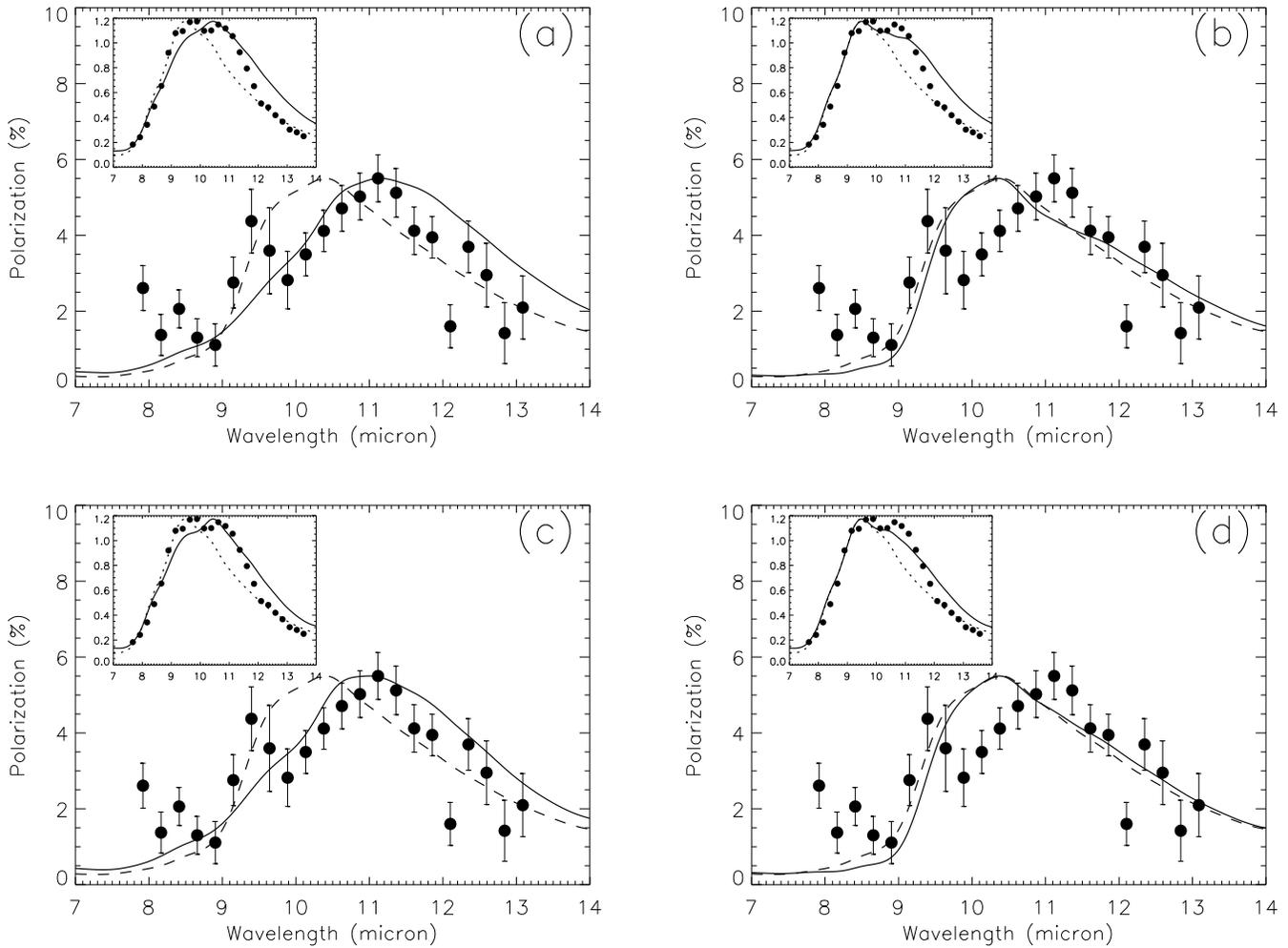}
\caption{SVS13 polarisation spectrum (solid circles)
overlaid with various polarisation models (solid lines)
calculated for spheroids in the Rayleigh approximation.
The data around 9.5~$\umu$m are compromised by telluric
ozone. Insets show the optical depth profile extracted
from the conventional flux density spectrum. The dashed
line in all panels (insets) corresponds to the
polarisation (absorption) expected from `astronomical
silicate' of \citet{Draine2003sbi}. In (b) and (d) are
models with a mantle of amorphous (b) and crystalline (d)
water ice, and a mantle-to-core volume ratio of 0.5.
Models in (a) and (c) are for silicate-ice mixtures, again
using amorphous (a) and crystalline (c) ice with a volume
fraction of 0.375 and a Maxwell-Garnett mixing rule for
the dielectric function of the effective medium. In all
cases the grain cores are oblate with an axial ratio of
2:1.
}
\label{svs13-pol-fig1}
\end{figure*}

\begin{figure*}
\includegraphics[width=1.0\textwidth, angle=180]{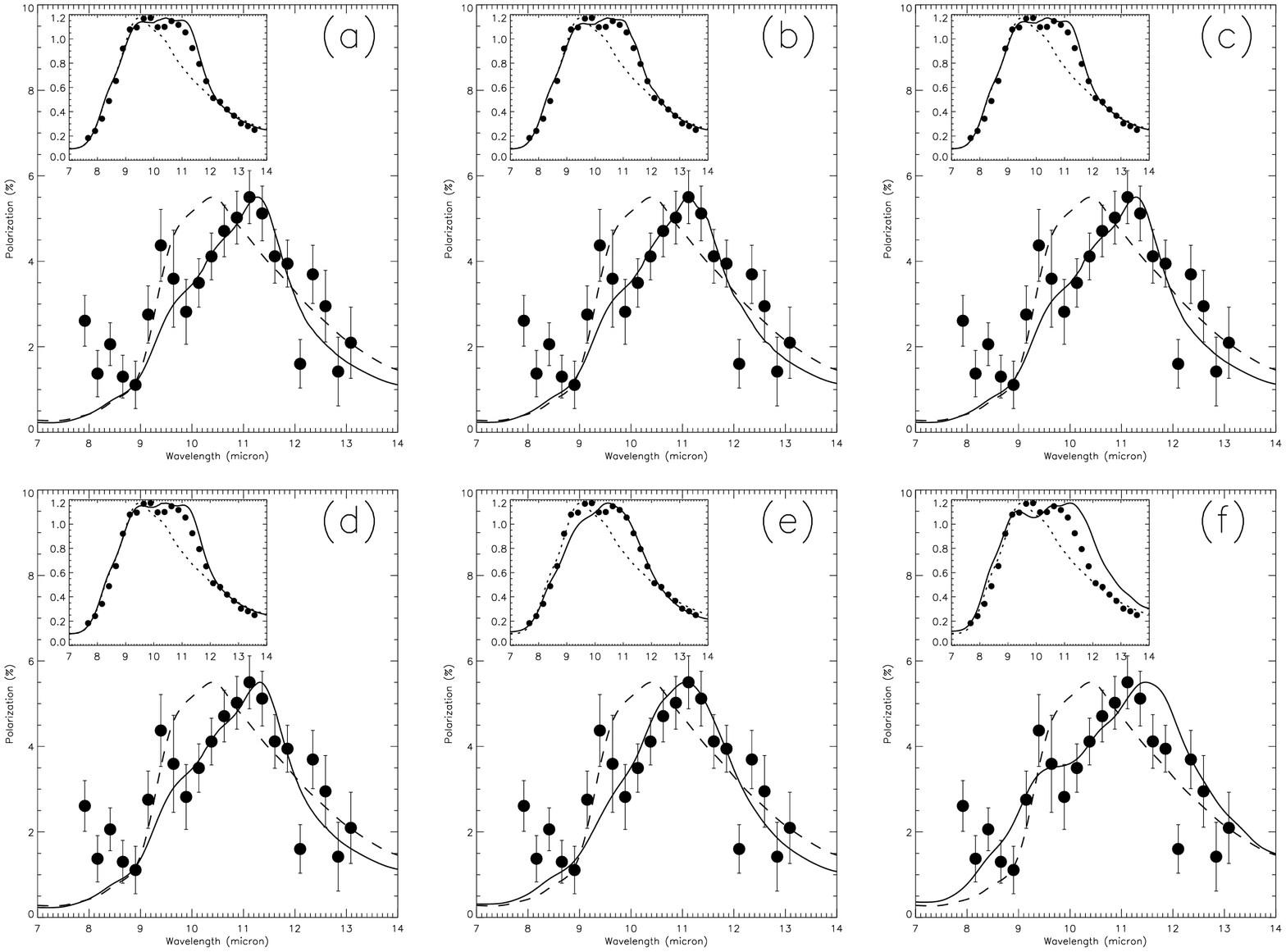}
\caption{SVS13 polarisation spectrum (solid circles)
overlaid with various polarisation models (solid lines)
calculated for spheroids in the Rayleigh approximation.
The data around 9.5~$\umu$m are compromised by telluric
ozone. Insets show the optical depth profile extracted
from the conventional flux density spectrum. The dashed
line in all panels (insets) corresponds to the
polarisation (absorption) expected from the
\citet{Draine2003sbi} `astronomical silicate'. Panels
(a)$-$(c) use SiC optical data from \citet{Pitman2008opo}
for cubic (a), hexagonal e-ray (b) and hexagonal o-ray (c)
samples. Panel (d) uses $\alpha$-SiC refractive indices
from \citet{Choyke1985scs}, panel (e) `astronomical SiC'
from \citet{Laor1993sco} and panel (f) $\alpha$-SiC from
\citet{Pegourie1988opo}. In all cases the grains are
oblate with an axial ratio of 2:1.
}
\label{svs13-pol-fig2}
\end{figure*}

Similar to the conventional spectrum, the MIR polarisation
spectrum of SVS13, first discussed in
\citet{Wright1999msa} and \citet{Smith2000sim}, is perhaps
the most unique ever observed (see
Figures~\ref{svs13-pol-fig1} and \ref{svs13-pol-fig2}).
Given the peak occurs beyond 11~$\umu$m and the overall
profile has an approximate `tilde' shape, properties
similar to those predicted for polarised emission
(\citealt{Martin1975sio}; \citealt{Aitken1989saa}), at
first glance it is not even obvious whether the spectrum
is representative of dichroic absorption or instead
polarised emission. Also, that the total flux density
spectrum is obviously dominated by absorption does not
necessarily imply the polarisation profile would also be
indicative of absorption. Instances can occur where the
polarisation is dominated by emission but the conventional
flux density shows deep absorption, an effect first
predicted by \citet{Thronson1979opb} and seen in a few
sources (e.g., NGC7538~IRS1 and RCW57~IRS1 in
\citealt{Smith2000sim}).

Fortunately in the case of SVS13 there is additional
polarisation data in the NIR, up to almost 3.5~$\umu$m,
which can be compared with the MIR
(\citealt{Holloway2002sot} and references therein).
Most specifically the measured polarisation position
angles in the two spectral regions are almost identical at
$\sim$$50 \pm 5$~degrees. The NIR polarisation is almost
certainly due to dichroic absorption since it rises
through the 3-$\umu$m water-ice band. There can be no
polarised emission through this band because the ice would
obviously melt at the 500$-$1,000-K temperatures
characteristic of NIR emission (and it is unclear if
relevant grain alignment mechanisms could operate at such
a high temperature). Thus, given the equality of the
position angles, it is very likely that the MIR
polarisation is also due to dichroic absorption.

Having established the polarisation process as dichroic
absorption it remains to determine what type of dust can
produce such a unique 8$-$13-$\umu$m profile, and
especially a peak beyond 11~$\umu$m. Possible candidate
carriers with bands around 11~$\umu$m and longer, inferred
in other astronomical spectra and/or found within
pre-Solar meteoritic grains, include crystalline silicates
and SiC, as shown in Figure~\ref{all-comp}, as well as
water ice, Polycyclic Aromatic Hydrocarbons (PAHs),
aluminium oxides and carbonates. Polarisation models have
been run for all of these potential constituents, either
as inclusions mixed in with an amorphous silicate host
matrix or as a confocal mantle coating an amorphous
silicate core.

Details of the modelling will be presented elsewhere
(Wright et al.\ in preparation), but are based on the
Rayleigh approximation for spheroidal particles. At MIR
wavelengths, according to \citet{Somsikov1999ota}, this is
adequate for grain radii (of an equivalent volume sphere)
up to at least 0.5~$\umu$m, where the relative difference
between exact and Rayleigh solutions is less than
10~per~cent for the water ice coated grains of their work.
Relevant formulae for the absorption and polarisation
cross sections of both bare and mantled spheroids are
given in \citet{Draine1984opo} and \citet{Lee1985iea},
whilst those for an `average' dielectric function of a
mixture, using effective medium theory such as the
Maxwell-Garnett rule, are provided in
\citet{Bohren1983aas}. In all cases we have used an oblate
core with principal axis ratio of 2, parameters which are
reasonably constrained from other observations (e.g.,
\citealt{Hildebrand1995tsa}; \citealt{Draine2006xsb}).

The amorphous silicate optical constants are those of
`astronomical silicate', first formulated by
\citet{Draine1984opo}, and subsequently updated by
\citet{Laor1993sco}, \citet{Weingartner2001dgd} and
\citet{Draine2003sbi}. The MIR component of these optical
data sets was constructed to reproduce the spectrum of
Trapezium, so our choice is consistent with the emissivity
function we used to model the conventional (total flux
density) spectrum in \S\ref{model}.

As might be expected based on the absorption coefficients
shown in Figure~\ref{all-comp}, polarisation features of
crystalline silicates, such as inclusions of the
crystalline olivine of \citet{Mukai1990oco}, are simply
too narrow to match the observed SVS13 polarisation
spectrum. They instead produce a spectrum which looks much
like the polarisation profile of the massive embedded
young stellar objects (YSOs) AFGL2591
(\citealt{Aitken1988iso}; \citealt{Wright1999msa}) and
IRAS13481$-$6124 \citep{Wright2008amp}, two sources with a
sharp polarisation feature at $\sim$11.2~$\umu$m
superposed on the much broader amorphous silicate band.
Similar comments hold for PAHs and carbonates. Whilst its
resonance is relatively broad, aluminium oxide inclusions
merely produce a shoulder, or long-wavelength wing, on the
amorphous silicate polarisation profile
\citep{Wright2002tma}.

The long-sought but rarely identified 12-$\umu$m
librational band of water ice, occurring between $\sim$12
and 13~$\umu$m for the crystalline and amorphous phases
respectively \citep{Maldoni1998aso}, may be thought to be
a strong candidate. SVS13 does have the characteristic
3.1-$\umu$m H$_{2}$O ice absorption band, which
\citet{Parise2003sfs} note is more like that from
crystalline than amorphous ice. Its optical depth however
is not particularly high, only around 0.55. This compares
to values of around 1 to 3.5 in a sample of embedded YSOs
in \citet*{Smith1989afi}. Their study included another
source with an unusual 8$-$13-$\umu$m spectrum, namely the
eastern component of the double source AFGL961, with a
3.1-$\umu$m optical depth of 2.46. 

Indeed, the only previous identification of the
librational band was made for AFGL961, firstly by
\citet{Cox1989tlo} using {\it IRAS} and then by
\citet{Smith2011tlb} for each component using
{\it Spitzer}. This has recently been questioned by
\citet*{Robinson2012twl} on the basis of radiative
transfer models suggesting the 8$-$13-$\umu$m silicate
feature could be self-absorbed, leading to a local minimum
between the 10- and 20-$\umu$m silicate bands occurring
around 13~$\umu$m and thus mimicking ice absorption.
However, their models only do an average job of
reproducing the overall shape of the observed
2$-$20-$\umu$m spectrum and profile of the 8$-$13-$\umu$m
feature. Also, they only modelled AFGL961E and not
AFGL961W, which has a prominent silicate emission feature
but still shows the same apparent absorption at 13~$\umu$m
(as well as CO$_{2}$ ice absorption at 15~$\umu$m).
Whatever the case, as presented in \citet{Boogert2008tcs},
there is sufficient evidence that the librational band has
been detected toward at least a few YSOs (although none
look much like SVS13 itself).

Thus, polarisation models with water ice either as a
mantle or as inclusions were attempted. Ice existing as a
mantle on silicate cores is the `standard' model for dust
in molecular clouds (e.g., \citealt{Gibb2004iit}). On the
other hand, the idea behind ice as inclusions is that if
silicate grains are initially porous or fluffy, then
within a cold molecular cloud (or disc) water ice may
condense to fill the pores. Alternatively, ice-mantled
grains may collide and stick together, resulting in a
similar type of grain structure. 

Whatever the case, core-mantle grains cannot produce the
position of the polarisation peak in SVS13. See
Figure~\ref{svs13-pol-fig1}, which shows in (b) and (d)
models using a mantle of respectively amorphous and
crystalline water ice (optical constants from
\citealt{Leger1983poa} and \citealt*{Bertie1969aoi},
respectively). The mantle-to-core volume ratio is 0.5, but
in fact mantles do not sufficiently change the overall
polarisation profile even up to mantle-to-core volume
ratios of 4 or more. The 10.3-$\umu$m peak of bare
amorphous silicates is still clear, and indeed sharper and
higher than the water ice peak around 12~$\umu$m, whilst
the observed SVS13 peak lies in the trough between them. 

On the other hand, as seen in
Figure~\ref{svs13-pol-fig1}~(a) and (c), a silicate-ice
mixture can reasonably match at least the overall shape
and peak position of the SVS13 polarisation spectrum for
an ice-to-silicate volume ratio of 0.375. A serious flaw
however is that it predicts large polarisation in the
3-$\umu$m water ice band, e.g., an excess of
$\sim$6$-$8~per~cent over the continuum compared to an
observed value of $\leq$~1.5~per~cent described in
\citet{Aitken1996sos} and \citet{Chrysostomou19963ms}.
This assumes that grains are small enough that the
Rayleigh approximation remains valid at 3~$\umu$m,
reasonable for grain sizes $\leq$~0.5~$\umu$m. See tab.2
in \citet{Somsikov1999ota}, where the relative difference
between exact and Rayleigh solutions only increases to
10$-$15~per~cent at 3~$\umu$m for either oblate or prolate
grains with an `equivalent volume' spherical radius of
0.5~$\umu$m.

Included in each panel of Figure~\ref{svs13-pol-fig1} is
an inset showing the optical depth extracted from the UCLS
spectrum of SVS13 as presented in \citet{Smith2000sim},
compared to the absorption cross section of the respective
models. The optical depth was calculated by fitting a
Planck function between the short and long wavelength ends
of the spectrum, after correcting them for a grey
component of emissivity taken to be the Trapezium
emissivities at 7.5 and 13.5~$\umu$m. The inferred colour
temperature is around 350~K, not too dissimilar to the
model-based temperatures listed in
Table~\ref{fit-results}. In no case does water ice, either
as a mantle or as inclusions, match the optical depth
profile. We therefore conclude that water ice is highly
unlikely to be responsible for the unique SVS13 MIR
spectrum, unless its optical properties are much different
to those used here (which are extremely similar to those
presented in many other publications).

Thus, the only remaining alternative is SiC and indeed
this provides the best match to the observed SVS13 MIR
polarisation spectrum. See Figure~\ref{svs13-pol-fig2}
which shows polarisation models of SiC inclusions within
an amorphous silicate matrix, using the Maxwell-Garnett
mixing rule to calculate the effective dielectric
function. Several different dielectric functions of SiC
were trialed, namely: cubic (3C or $\beta$) and two
samples of hexagonal (6H or $\alpha$) for E both
perpendicular and parallel to the c-axis, derived by
\citet{Pitman2008opo} from single-crystal reflectance
spectroscopy; $\alpha$-SiC in \citet{Choyke1985scs}, also
from reflectivity measurements; an `astronomical SiC' from
\citet{Laor1993sco} derived from several different
laboratory data sets; $\alpha$-SiC determined from
particle transmission measurements and presented by
\citet{Pegourie1988opo}.

All can provide good matches to the overall profile of
the observed spectrum, with respective volume fractions of
the SiC inclusions of 0.1 for the
Pitman\nocite{Pitman2008opo} et al.\ and Choyke \&
Palik\nocite{Choyke1985scs} data sets, and 0.25 for the
Laor \& Draine\nocite{Laor1993sco} and
Pegourie\nocite{Pegourie1988opo} optical data. Whilst the
models are all qualitatively similar, apart perhaps from
the Pegourie case, the best match appears to be for the 6H
(hexagonal) e-ray optical data of Pitman et al.\ (2008;
Figure~\ref{svs13-pol-fig2}~b)\nocite{Pitman2008opo}.
Specifically, the wavelength of peak polarisation and the
profile shapes of both the polarisation and optical depth
more closely follow the observations.

That the SiC polarisation signature dominates over that of
silicates is testament to its stronger refractive indices
and band strength, i.e., absorption cross section per unit
volume. In at least the crystalline samples this is
accentuated by a polarisation reversal across the strong
resonance (\citealt{Martin1975sio};
\citealt{Hong1978prb}), which essentially subtracts
polarisation from the amorphous silicate at 10~$\umu$m.

In the interests of completeness, we computed models
with SiC as a confocal mantle on a silicate core.
These did not provide anywhere near as good a match to the
observed optical depth and polarisation profiles.
Similarly, models using the optical data of amorphous thin
films of SiC from \citet{Mutschke1999ipo} and
\citet{Larruquert2011soc} did not match the spectra, being
either too broad and/or peaking at the wrong wavelength.
Given their weaker refractive indices than for the
crystalline phases they also necessitated a larger
relative abundance of SiC (compared to silicates).
Finally, we replaced the `astronomical silicate' with a
magnesium-iron laboratory olivine,
Mg$_{0.8}$Fe$_{1.2}$SiO$_{4}$, from
\citet{Dorschner1995sti}, motivated by its good match to
the diffuse ISM MIR polarisation profile presented in
\citet{Wright2002tma} and \citet{Wright2005msd}. This
provided very similar fits to the data.

In summary, both the conventional and polarisation
spectra require the presence of a material with optical
properties much like SiC, inclusive of a high band
strength, a central wavelength near 11.3~$\umu$m and full
width of a few microns.  We thus assess the carrier of the
additional absorption from approximately 10 to 12~$\umu$m
in SVS13 to be SiC itself. But given the variation in the
optical properties of specific types of SiC found in the
literature a reliable estimate of its abundance is
difficult. This is apart from a general statement that it
is higher than in any other young star envelope or disc of
which we know.

\subsubsection{Crystalline silicates}\label{cry.sil}

The {\em ISO} crystalline silicate revolution
\citep{Jaeger1998sti} came about
because the {\em ISO} satellite was equipped with
instruments capable of making measurements at longer
wavelengths, where crystalline silicate features are not
overwhelmed by those of more abundant and warmer amorphous
silicates, as is usually the case in the $N$-band
\citep{Molster2010tmo}.
There are three plausible mechanisms adept at
producing crystalline silicates.
\begin{enumerate}
\renewcommand{\theenumi}{\arabic{enumi}.}
\item {\em Evaporation and re-condensation} Under high
temperature and pressure condition in the immediate
vicinity of the central star ($<$~1~au), dust grains
can be evaporated and then re-condense as crystals
(e.g., \citealt{Grossman1972cit}).
\item {\em Thermal annealing} There have been at
least three probable heat sources suggested in
circumstellar discs
[accretion luminosity in the inner region ($\sim$1~au),
\citet{Gail2004rmi}; shock heating at a few au,
\citet{Harker2002aos}; and disc surface layer annealing
during accretion outburst, \citet{Abraham2009efo}].
Whatever the actual heat source(s) may be, the essence
of the mechanism is simple annealing (i.e., heating of
amorphous silicates and subsequent cooling to re-order the
lattice structure).
\item {\em Low-temperature crystallisation} Put forward
by Yamamoto and co-workers (e.g.,
\citealt{Yamamoto2005amo}),
provided that an amorphous silicate core is coated with an
organic (carbonaceous) refractory layer
\citep*{Kimura2003csa}, a chemical reaction can be
triggered at a low temperature (a few hundred K) to
re-arrange the silicate lattice structure.
\end{enumerate}

It is interesting to note that, so far, most detections
of crystalline silicate features towards young stars have
been in emission (e.g., \citealt{Watson2009csa};
\citealt{Olofsson2009css}; \citealt{Juhasz2010dei}), with
only one recent report of clear detection of absorption
\citep{Poteet2011asi} to date,\footnote{A tentative
identification has been made for deeply embedded low-mass
protostars IRAS~03445+3242 and HH46~IRS
\citep{Boogert2004sst}, and for the high-mass YSO
AFGL2591 \citep{Aitken1988iso}.
Crystalline silicate absorption features towards other
types of objects, e.g., ultra-luminous infrared galaxies
\citep{Spoon2006tdo}, have also been reported.}
a fact that may or may not favour one crystallisation
mechanism over the others.
Inclusions of the two most abundant crystalline silicate
species, forsterite (Mg$_{2}$SiO$_{4}$) and enstatite
(MgSiO$_{3}$), in absorption, substantially improve
the model fitting in the current study.
Equilibrium calculations suitable for the inner disc
regions ($\lesssim$~1~au, \citealt{Gail2004rmi}) show that
these magnesium-rich crystalline silicates should
dominate.

However, the crystalline silicate features are observed
in absorption here and therefore the dust grains
responsible for the signatures are likely to reside in the
cold, outer regions. Apart from the low-temperature
crystallisation described above, some means of heating is
required to turn the amorphous lattice structure into
crystalline, and if heating is only possible in the
vicinity of the central star, an effective
transport/mixing mechanism is required for crystalline
silicates to be found in the remote, cooler regions.
Indeed, crystalline silicates appear to be concentrated in
the centre of many T-Tau systems, with more than
90~per~cent of the 84 objects observed by
\citet{Watson2009csa} showing strong 10-$\umu$m
crystalline silicate features but only about 50~per~cent
having detectable 20$-$38-$\umu$m signatures, most
probably indicating the temperature range of the
crystalline silicates ($T \sim 200$~K at $\la$~a few~au,
\citealt{Watson2009csa}).
\citet{Olofsson2009css}, who, as part of the {\em Spitzer}
cores-to-discs legacy programme, observed 108 young stars
(66 of which were known T-Tau stars), identified what
they termed the `crystallinity paradox', wherein they
found the cold crystalline features at wavelengths
$>$~20~$\umu$m more frequently than those at
$\sim$10~$\umu$m. However, when they re-examined it by
analysing a further 58 {\em Spitzer} IRS spectra, they
found a simultaneous enhancement of crystallinity in both
the inner, warm and outer, cold regions
\citep{Olofsson2010css}.
\citet{Sargent2009dpa} also reported comparable
crystalline silicate abundances in the inner and outer
disc regions in a {\em Spitzer} IRS study of 65 T-Tau
stars.
Although details may differ slightly, an undisputed
universal outcome of all these investigations is the
existence of processed (i.e., crystalline) materials in
the outer, cooler parts of the circumstellar environment,
as well as in the warmer, inner regions in the vicinity of
the central star.

Comets contain crystalline silicates (e.g., Comet
Hale-Bopp, \citealt{Crovisier1996tis};
1997\nocite{Crovisier1997tso}). The most definitive
confirmation was once again found in the Stardust sample
return from Comet 81P/Wild~2, the crystalline silicate
mass fraction of which was as high as $\sim$0.5$-$0.65
\citep{Westphal2009mfo}.
The 81P/Wild~2 specimens also included chondrule-like
particles \citep{Nakamura2008coi}, a clear
indication that materials that constitute the comet
experienced high temperatures in the past.
Another clue to the apparent heating episode(s) in the
Stardust samples is the presence of Calcium-,
Aluminium-rich Inclusion (CAI)-like grains
\citep{Zolensky2006map}. However, as mentioned in
\S\ref{sic.sec}, 81P/Wild~2 belongs to the short-period
JFCs, which are believed to originate in the Kuiper Belt,
extending from the orbit of Neptune (at $\sim$30~au) to
approximately 50~au, in which the surface temperature of
the most famous of its resident, Pluto, is only
$\sim$40~K \citep*{Stern1993efa}.

To overcome these facts at odds with each other (existence
of high-temperature phases in cold regions), a number of
mechanisms have been proposed to mix/transport radially
the dust grains transformed near the central star. These
include: the bipolar outflow `X-wind' model (e.g.,
\citealt*{Shu1996taa}), accretion disc turbulence (e.g.,
\citealt{Gail2001rmi}; \citealt{Bockelee-Morvan2002trm}),
spiral arms in marginally gravitationally unstable
discs (e.g., \citealt{Boss2004eot}), and outward transport
in the disc mid-plane (e.g., \citealt{Keller2004rmi};
\citealt{Ciesla2007oto}).
Now, both chondrules and CAIs are sizable entities
that often measure hundreds of $\umu$m and some as large
as mm (chondrules) or even cm (CAIs) in diameter
(e.g., \citealt{Scott2005cma};
\citealt{MacPherson2005cim}). Although particles that
large in the 81P/Wild~2 capture are rare
(\citealt{Horz2006ifo}; \citealt{Burchell2008coc}),
analogous to chondritic meteorites that contain
large chondrules and CAIs, a similar size distribution
may also be expected to exist in their probable natal
bodies such as comets \citep{Meier2014acc}.
A recent modelling effort \citep{Hughes2010pti} has found
that particles of the size of chondrules and CAIs (a few
mm or larger) cannot easily be transported out to the
comet-forming region, regardless of their initial
location. If this is the case, an in-situ production of
melted grains\footnote{Note that grains in question here
are not crystalline silicates so the low-temperature
crystallisation of Yamamoto and co-workers mentioned
earlier would not apply.} may have to be invoked. We will
come back to this point in \S\ref{cry.mech}.

\subsubsection{Annealed SiO$_{2}$}\label{ann.sio2}

Aforementioned equilibrium calculations also show that,
when the Mg/Si abundance ratio is $\lesssim$~1 (or
perhaps after a significant amount of Mg has been taken
up by forming forsterite), SiO$_{2}$ (and enstatite)
becomes a major condensate \citep{Ferrarotti2001mfi}.
\citet{Fabian2000sti} found that when they annealed
enstatite (MgSiO$_{3}$) smoke at 1,000~K for 30~hours, the
end product consisted of crystalline forsterite
(Mg$_{2}$SiO$_{4}$), tridymite (crystalline SiO$_{2}$) and
amorphous silica (SiO$_{2}$). This is a case similar to
that reported by \citet{Bowen1914tbs}, who cooled a
mixture of composition MgSiO$_{3}$ from the liquid state
and found that, first forsterite separated out and, after
further cooling, a mixture of silica and enstatite
condensed.
As crystalline silicates also exist, it may be deduced
that some form of heating has melted MgSiO$_{3}$ and
decomposed it into Mg$_{2}$SiO$_{4}$ and SiO$_{2}$.
We further speculate that, once again, taking the presence
of crystalline forsterite and enstatite as evidence, the
cooling rate is just so that it would allow re-ordering of
lattice structures to produce annealed SiO$_{2}$. Or, that
crystalline SiO$_{2}$ is dominant over its amorphous
counterpart in the stages immediately after their
formation.

As mentioned earlier, SiO$_{2}$ has been identified in
the MIR spectra of some young stars.
\citet{Honda2003doc} observed a T-Tauri star
Hen~3-600A in the $N$-band and their best-fit model
included $\alpha$-quartz (i.e., a type of
crystalline SiO$_{2}$, from \citealt{Spitzer1961ilb}).
\citet{Sargent2006dpi} used optical constants of
$\alpha$-quartz from \citet{Wenrich1996oco}
to analyse $N$-band spectra of 12 T-Tau stars. They
found a small amount (up to a few per cent by mass) of
crystalline SiO$_{2}$ in about a half of the objects
studied. Amorphous silica \citep{Henning1997lip}
and silica-rich glass \citep{Koike1989oco} emissivities
were ruled out by \citet{Sargent2009sip}, this time
observing 5 more, different T-Tau stars, but inclusion of
annealed SiO$_{2}$ \citep{Fabian2000sti} gave a better
fit. A {\em Spitzer} IRS spectrum of a $\beta$-Pictoris
analogous (i.e., debris-disc) star HD172555, which
suggested the presence of amorphous silica and SiO gas,
was examined by \citet{Lisse2009acs}. The 12.6-$\umu$m
feature that would have indicated the existence of
crystalline silica was not detected. They proposed
planetesimal-scale ($\sim$km-size) hyper-velocity
($>$~10~km~s$^{-1}$) impacts as the mechanism that
produced fine (amorphous) silica dust and SiO gas.
More recently, \citet{Fujiwara2012sbd} obtained a
{\em Spitzer} IRS spectrum of another debris-disc star
HD15407A and concluded that incorporating an almost equal
amount of fused (i.e., amorphous) quartz
\citep{Koike1989oco} and annealed silica
\citep{Fabian2000sti} provided the most satisfactory fit.
They also favoured a similar scenario to that adopted by
\citet{Lisse2009acs} for HD172555.

Although the number of samples is rather limited and most
studies mentioned above did not set out specifically to
probe crystalline or otherwise of silica dust structure,
overall it is tempting to conclude that younger stars
(T-Tauri stars, $\la$~10~Myr) are more likely to possess
more crystalline than amorphous SiO$_{2}$, while with
slightly more evolved sources (debris-disc bearing stars,
$\ga$~10~Myr), the balance is tipped in the other
direction. Also, if hyper-velocity planetesimal impacts
are indeed required for the production of amorphous
silica, this alone points to the latter stages in the
evolution of protoplanetary discs. If this scenario is
correct, then the natural implication may be that the
circumstellar environment of (still deeply embedded and
therefore much younger) PMS star SVS13 should contain
mostly crystalline SiO$_{2}$ but not much (if at all) of
its amorphous sibling.

\subsection{Possible crystallisation mechanism and planet
formation}\label{cry.mech}

We have demonstrated that
crystalline silicates (forsterite and enstatite) and
annealed SiO$_{2}$ exist in the cold, outer regions of
circumstellar environment of the PMS star SVS13. Although
we did not directly detect larger chondrules and CAIs in
the current study, they may exist if the typical dust size
distribution in chondritic meteorites also holds in the
remote, cooler regions (at least in the comet-forming
region at a few tens of au from the central star).
If that is the case, an in-situ transformation mechanism
of dust grains is probably required.

The runaway growth of planetesimals is the `standard'
model for terrestrial planet formation (e.g.,
\citealt{Raymond2014tpf}). Outer gas giants, like Jupiter
and Saturn, can be formed in a similar fashion via what is
called the core-accretion (CA) model (e.g.,
\citealt{Helled2014gpf}). But also feasible is if the
remote, cold regions of the protoplanetary disc become
gravitationally unstable and fragment, forming
self-gravitating clumps, some of which will ultimately
become gas giants (see e.g., \citealt{Helled2014gpf} and
references therein).

According to the latter mechanism, commonly referred to as
the disc instability (DI) model, fragmentation can
take place at $\ga$~50~au and clumps ranging from a few to
$\sim$10~$M_{J}$ (Jupiter masses) may form
(e.g., \citealt{Boley2010cit}; \citealt{Forgan2011tjm};
\citealt{Rogers2012tfo}). As these
massive clumps contract and migrate inwards, the internal
temperature rises and it can reach $\ga$~1,000~K, sufficient
to produce crystalline silicates, chondrules and CAIs
(and dissolving amorphous carbon coating on SiC grains)
{\em in~situ}.
When they are tidally disrupted, these
thermally altered particles may be distributed at a wide
range of radii to be re-captured by, e.g., comets
(\citealt{Vorobyov2011dom}; \citealt*{Nayakshin2011ttd}).

Of course, thermal processing of dust species can still
occur in the immediate vicinity of the central star and
the dust grains may be transported out to the
comet-forming region of the protoplanetary disc. One of
the plausible dust heating processes that could produce
chondrules and CAIs (and at the same time a possible
transport/mixing mechanism), is the X-wind model of
Shu et al.\ (i.e., jets near the central star).
But this has been criticised by \citet{Desch2010ace}, who
prefer shock-heating, due to DI but at a few au (e.g.,
\citealt{Desch2002amo}; see also \citealt{Desch2012tio}).
However, as mentioned in \S\ref{cry.sil}, it is
difficult, if not impossible, to move large particles such
as chondrules and CAIs out to the comet-forming region,
wherever their initial location may have been
\citep{Hughes2010pti}.
Even if shocks can be triggered, the gas density at the
distance of the Kuiper Belt ($\ga$~30~au) is such that
chondrule formation by shock-heating might not be
feasible.
\citet*{Iida2001ash} estimate the pre-shock density of at
least $10^{14.5}$~cm$^{-3}$ is required, while both the
minimum-mass Solar Nebula model of \citet{Hayashi1981sot}
and the minimum-mass extra-Solar nebula of
\citet{Chiang2013tme} predict a much lower density of the
order $\sim$10$^{10-11}$~cm$^{-3}$ at 30~au.

We recall that SVS13 is a close binary, with a separation
of $\sim$0.3~arcsec~$\approx$~71~au. Theory and model
simulations depict some common features in the early
stages of binary evolution: three separate discs
(individual circumstellar discs around each component of
the binary system and a circumbinary disc) and
a clearing between the two classes of discs (e.g.,
\citealt{Artymowicz1994dob}; \citealt{Bate1997adb};
\citealt{Gunther2002cbe}). Under certain conditions,
however, one binary component (primary) will gain a large
circumstellar disc, while the other (secondary) only a
small one (or none at all), with little to no
circumbinary disc \citep{Bate1997adb}. The primary
is usually the dominant dust continuum source
in PMS binary/multiple systems (e.g.,
\citealt{Harris2012arc}), and apart from a few outstanding
cases, circumbinary discs appear to be elusive
(\citealt{Monin2007dei}; \citealt{Harris2012arc}).
In the case of SVS13, only one member of the binary
(VLA~4B) exhibits significant dust emission
\citep{Anglada2004asc} and just a single object has been
found in the NIR (which \citealt{Hodapp2014tlr} identify
as VLA~4B) as we do here in the MIR.
Furthermore, the unresolved radio source exhibits a larger
flux density than the sum of the resolved individual
binary components, implying the existence of an extended
envelope \citep{Anglada2004asc}.
Also commonly seen in snapshots from simulations are
spiral arms in all three discs due to the tidal
interactions of the central binary. Whether such
gravitational interplay and structures inhibit or enhance
DI-induced fragmentation is a matter of debate
\citep*{Mayer2010gii}.

The `clearing' or a gap between the circumbinary and the
central circumstellar discs is also a product of the tidal
interactions between the binary components, as the
protoplanetary discs are stripped of materials and
truncated to substantially smaller sizes compared to those
around single stars (e.g., \citealt{Artymowicz1994dob}).
Indeed, \citet{Anglada2004asc} found a compact structure
at the position of VLA~4B of radius $\sim$30~au, clearly
smaller than that often measured for discs around single
stars of $\ga 100$~au (see e.g., \citealt{Williams2011pda}
and references therein). Discs are not only truncated but,
in such a dynamic environment, they are also rapidly
dispersed within a short period of time ($\la 1$~Myr),
considerably shorter than a characteristic disc lifetime
of $\ga 2.5$~Myr normally expected for single stars (e.g.,
\citealt{Mamajek2009ico}), though the effect could be
slightly milder for a binary separation of $\sim$70~au for
SVS13 (\citealt{Monin2007dei}; \citealt{Cieza2009pcd}).
Although this type of disruptive impact might be thought
to affect the planet formation negatively, Bonavita \&
Desidera (2007\nocite{Bonavita2007tfo}; see also
\citealt*{Bonavita2010tfo}) concluded that the detection
frequencies of planets in single and binary systems are
statistically identical.
Most interestingly, \citet{Duchene2010pfi}
showed that, while binaries with separation wider than
100~au are indistinguishable from single stars in terms of
trends regarding their protoplanetary discs, debris discs,
and planets, close binaries in the range from 5 to 100~au
exclusively host giant planets (i.e., those with
mass $\ga 1$~$M_{J}$), or exhibit a distinct lack of
planets smaller than 1~$M_{J}$.
The two modes of planet formation discussed earlier
present notably different timescales: CA with $\ga$~a few
Myr and DI $\la$~1~Myr \citep{Helled2014gpf}.
Considering the short lifetime of circumstellar discs in
close binaries, it could be the case that DI is the only
effective means of forming planets in them, and if so, the
finding of \citet{Duchene2010pfi} might just be a natural
consequence of this.

Recent observations of the young low-mass binary system
GG~Tau~A \citep{Beck2012cga} have revealed that a
`streamer' of dust and gas, originating from the
circumbinary disc and feeding the inner circumstellar
discs can shock-excite H$_{2}$ gas at the interface, where
the gas temperature was found to be in excess of 1,500~K.
Unlike the shock-heating of particles (due to DI) rejected
earlier because of the low density in the outer parts of
the circumstellar disc, we surmise that a stream of
material from the circumbinary disc piling up at the
outer rim of the protoplanetary disc, combined with a
higher shock velocity generated by streamers
(20$-$30~km~s$^{-1}$ for streamers vs.\ $<$~10~km~s$^{-1}$
for DI, \citealt{Desch2002amo}; \citealt{Harker2002aos};
\citealt{Beck2012cga}), could make this a
viable mechanism for melting/annealing dust
grains {\em in~situ} in the remote regions of
circumstellar discs.

We therefore suggest that, amongst the processes that are
capable of thermally altering dust grains, an in-situ
mechanism is favoured for SVS13. Furthermore, the DI model
for giant planet formation -- provided that fragmentation
occurs in circumstellar discs in binary systems -- along
with the shock-heating mechanism detected by
\citet{Beck2012cga}, is the leading candidate for
producing crystalline silicates (and chondrules \& CAIs),
annealing SiO$_{2}$, as well as possibly dissolving the
feature-concealing amorphous carbon layer from the SiC
core.
It remains to be seen if the scenarios just presented are
rare or only occur in a short-lived evolutionary phase in
shaping the distinctive $N$-band spectrum of SVS13. To
this end, we are embarking on the search for similar
objects and at the same time, it is hoped that work both
in the laboratories and theoretical studies will advance
sufficiently in the near future to solve, for example, the
missing SiC problem.

\section{Conclusions}

The $N$-band spectrum of the low-mass PMS close binary
system SVS13 is presented. The unique and complex spectrum
is best modelled by a mixture of amorphous silicates,
crystalline forsterite, crystalline enstatite, annealed
SiO$_{2}$, and, most intriguingly of all, SiC, which has
never before been identified in the circumstellar
environment of a young star. The requirement for the
inclusion of silicon carbide is also affirmed from the MIR
polarimetry perspective. All these signatures are seen in
absorption, implying the dust particles reside in
remote, cold regions of the circumstellar environment.
Speculation is made on possible origins of the dust
species, especially those that have been processed and
altered thermally, namely crystalline silicates and
annealed SiO$_{2}$. The DI-induced fragmentation and the
subsequent contraction and disruption of the giant planet
embryos, along with the newly discovered shock-heating
mechanism at the interface between the circumbinary and
circumstellar discs, may be able to modify those dust
grains {\em in~situ} in the outer, cooler parts of the
circumstellar environment. The heating episode(s) provided
by one or both of these processes can also feasibly
reveal the hidden SiC feature.

\section*{Acknowledgements}

Part of this work was conducted while TF was a Visiting
Fellow at School of Physical, Environmental and
Mathematical Sciences, UNSW Canberra, Australia. TF would
like to thank their warm hospitality during his stay.
CMW acknowledges support from the Australian Research
Council Future Fellowship FT100100495. We would also like
to thank the referee, Dr.\ Angela Speck, for her valuable
comments.

\label{lastpage}

\end{document}